\documentclass[10pt,a4paper,onecolumn]{vupreprint}
\usepackage[bindingoffset=0.5cm,textheight=25cm,hdivide={2.5cm,*,3cm}, vdivide={*,24cm,*}]{geometry}
\usepackage{amsmath}
\usepackage{amsfonts}
\usepackage{amssymb}
 \usepackage[numbers,square,comma,sort&compress]{natbib}
 \usepackage{fancyhdr}
 \usepackage{fancybox}
 \usepackage{authblk}
 \usepackage{siunitx}
 \usepackage{booktabs}
 \usepackage{multirow}
 \usepackage{soul}
 \usepackage{microtype}
 \usepackage[pdftex]{graphicx}
 \usepackage[pdftex,bookmarksnumbered=true,breaklinks=true]{hyperref}
\usepackage{upgreek}
\usepackage{textcomp}
\newcommand{\eqnref}[1]{Equation~(\ref{#1})}
\newcommand{\eqnsref}[2]{Equations~(\ref{#1})--(\ref{#2})}
\newcommand{\figref}[1]{Figure~\ref{#1}}			
\newcommand{\tabref}[1]{Table~\ref{#1}}			
\newcommand{\secref}[1]{Section~\ref{#1}}		
\newcommand{\um}[1]{\SI{#1}{\micro\metre}}

\newcommand{\pa}{\partial}						
\newcommand{\diff}[2]{\frac{\pa #1}{\pa #2}}				

\newcommand{\qbounce}{{\it{q}}{\sc{Bounce}}}				
\newcommand{\cannex}{\textsc{Cannex}}
\renewcommand{\a}{\alpha}

\newcommand{\vare}{\varepsilon}

\renewcommand{\Xi}{\Xi}

\newcommand{\inv}[1]{\frac{1}{#1}}					
\graphicspath{{./}{./figures/}}

 \doi{10.3390/sym11030407}
 \pacs{quantum vacuum; Casimir pressure; axion; non-Newtonian gravity}
 \journalref{G. L. Klimchitskaya, V. M. Mostepanenko, R. I. P. Sedmik, and H. Abele}{Symmetry}{11(3)}{407}{2019}{Prospects for Searching Thermal Effects, Non-Newtonian Gravity and Axion-Like Particles:
Cannex Test of the Quantum Vacuum}
 \author{{Galina L. Klimchitskaya}$^{1,2,}${\footnote{Correspondence: g.klimchitskaya@gmail.com}} }
 \author{Vladimir M. Mostepanenko$^{1,2,3,}$\footnote{vmostepa@gmail.com} }
 \author{Ren\'{e} I. P. Sedmik$^{4,}$\footnote{rene.sedmik@tuwien.ac.at} }
 \author{Hartmut Abele$^{4,}$\footnote{hartmut.abele@tuwien.ac.at}}
 \affil{$^1$Central Astronomical Observatory at Pulkovo of the Russian Academy of Sciences, Saint Petersburg 196140, Russia;}
 \affil{$^2$Institute of Physics, Nanotechnology and Telecommunications, Peter the Great Saint Petersburg Polytechnic University, Saint Petersburg 195251, Russia}
 \affil{$^3$Kazan Federal University, Kazan, 420008, Russia}
 \affil{$^4$Technische Universit\"{a}t Wien, Atominstitut, Stadionallee 2, 1020 Vienna, Austria}
\hypersetup{pdfauthor={\theauthors},pdftitle={\thetitle}}
\title{Prospects for Searching Thermal Effects, Non-Newtonian Gravity and Axion-Like Particles:
Cannex Test of the Quantum Vacuum}
\begin{document}
\maketitle
\begin{abstract}
We consider the \cannex{} (Casimir And Non-Newtonian force EXperiment) test of the quantum vacuum
intended for measuring the gradient of the Casimir pressure between two flat parallel plates at large
separations and constraining parameters of the chameleon model of dark energy in cosmology.
A modification of the measurement scheme is proposed that allows simultaneous measurements
of both the Casimir pressure and its gradient in one experiment. It is shown that with several
improvements the \cannex{} test will be capable to strengthen the constraints on the parameters
of the Yukawa-type interaction by up to an order of magnitude over a wide interaction range. The
constraints on the coupling constants between nucleons and axion-like particles, which are considered as the most
probable constituents of dark matter, could also be strengthened over a region
of axion masses from 1 to 100~meV.
\end{abstract}
\section{Introduction}
\label{sec:intro}
Since the development of quantum field theory, it has been appreciated that the quantum vacuum is a fundamental type
of physical reality which potentially contains all varieties of elementary particles and their interactions. Although an infinitely large energy density of zero-point oscillations of quantum fields (the so-called{ {\it virtual particles}}) could be considered to be catastrophic \cite{Adler:1995}, convenient self-consistent procedures have been elaborated on how to make
 it equal to zero in the empty Minkowski space and take it into account when calculating the probabilities of arbitrary processes
in the framework of the Standard Model. In doing so, the quantum vacuum is usually responsible for some part of the
measured quantity (for instance, the Lamb shift), whereas the rest of it is determined by the ordinary (real) particles.

There is, however, one physical phenomenon, where the measured quantity is determined entirely by the quantum vacuum.
This is the Casimir effect arising in the quantization volumes restricted by some material boundaries or in cosmological
models with nontrivial topology \cite{Elizalde:2012,Elizalde:1994,Elizalde:2003,Bordag:2014}.  An essence of this effect is that although the vacuum energy in
restricted or topologically nontrivial volumes remains infinitely large, it becomes finite when subtracting
the vacuum energy of empty topologically trivial Minkowski space. The negative derivative of the obtained finite vacuum
energy with respect to the length parameter (either a separation between the boundary surfaces or a scale of the
topologically nontrivial manifold) results in the Casimir force, which generalizes the familiar van der Waals force in the case
of larger separations when one should take into consideration the effects of relativistic retardation \cite{Bordag:2014}.

Recently it was understood that the quantum vacuum may be responsible for
{an} impressive phenomenon in
nature, i.e., for acceleration of  expansion of the Universe~\cite{Frieman:2008}.~This can be explained by an impact of the
energy density of quantum vacuum (which is often referred to as {{\it dark energy}) }corresponding to a nonzero
renormalized value of the cosmological constant in the Einstein
equations of general relativity theory \cite{6a}.

Another subject is that at sufficiently short
separations between the boundary surfaces, vacuum forces are stronger than Newtonian gravitation.
In this case, they form a background for testing new physics, such as Yukawa-type corrections to Newton's law
 of gravitation arising due to exchange of light hypothetical scalar particles \cite{Fischbach:1999} or due to spontaneous
compactification of extra spatial dimensions at the low-energy compactification scale \cite{Antoniadis:1998}. Forces of this
kind would alter the energy eigenstates of a neutron in the gravity potential of the Earth,
and are searched for by a technique called Gravity Resonance Spectroscopy~\cite{Cronenberg:2018,Jenke:2014a,Jenke:2011}
by the \qbounce{} collaboration. On the background of Casimir forces, one could also search for axions and other axion-like particles \cite{Bezerra:2014,Bezerra:2014a,Bezerra:2014b,Bezerra:2014c,Klimchitskaya:2015a,Bezerra:2016,Klimchitskaya:2017,Klimchitskaya:2017a}
which are considered as hypothetical constituents of dark matter. It is remarkable that taken together, dark
energy and dark matter contribute for more than 95\% of the energy of the Universe, leaving less than 5\% to the forms of
energy we are presently capable to observe directly~\cite{Frieman:2008}.

Precise measurements of the Casimir force revealed a problem that the experimental data agree with
theoretical predictions of the fundamental Lifshitz theory only under the condition that in calculations one disregards
the relaxation properties of conduction electrons and the conductivity at a constant current for metallic and
dielectric boundary surfaces, respectively (see review in \cite{Bordag:2014,Klimchitskaya:2009a} and more modern experiments \cite{Banishev:2012,Banishev:2013a,Banishev:2013b,Bimonte:2016,Chang:2011,Banishev:2012a}).
Theoretically, it was shown that an inclusion of the relaxation properties of conduction electrons and the conductivity at a constant
current in computations results in a violation of the Nernst heat theorem for the Casimir entropy
 (see review in \cite{Bordag:2014,Klimchitskaya:2009a} and further results~\cite{Klimchitskaya:2015b,Klimchitskaya:2015c,Klimchitskaya:2017b,Klimchitskaya:2017c}).  Taking into account that both the relaxation properties
of conduction electrons for metals and the conductivity at a constant current for dielectrics are
well studied really existing phenomena, there must be profound physical reasons for disregarding them in calculations of the
Casimir force caused by the zero-point and thermal fluctuations of the electromagnetic field.

All precise experiments on measuring the Casimir interaction mentioned above have been performed in the sphere-plate
geometry at surface separations below a micrometer. In this paper, we consider the \cannex{} (Casimir And Non-Newtonian force
EXperiment) that was designed to test the quantum vacuum in the configuration of two parallel plates at separations up to
\SI{15}{\micro\meter} \cite{Almasi:2015a,Sedmik:2018}. In~addition to the already discussed possibility of testing the nature of dark energy \cite{Almasi:2015a},
we consider here the potentialities of this experiment for searching thermal effects in the Casimir force, stronger
constraints on Yukawa-type corrections to Newton's gravitational law, and on the coupling constants of axion-like
particles.
For this purpose, a modification in the measurement scheme is proposed, which allows simultaneous measurement of
both the Casimir pressure and its gradient.
It is shown that after making several improvements to the setup, the \cannex{} test would be capable of performing the first observation of thermal effects in the Casimir interaction
and to strengthen the presently available limits on Yukawa-type corrections to Newtonian gravity by up to a factor of 10. Stronger limits could also be obtained on the
coupling constants of axion-like particles to nucleons within a wide range of axion masses.

The paper is organized as follows. In \secref{sec:exp} we briefly describe the experimental setup with two parallel plates and possible improvements. \secref{sec:thermal} is devoted to the calculation of thermal
effects in the Casimir pressure and its gradient at separations relevant for \cannex{}. In \secref{sec:constraints} the prospective constraints
on non-Newtonian gravity and axion-like particles, which could be obtained from the improved setup, are found.
\secref{sec:conclusion} contains our conclusions and discussion.

\section{Experimental Setup with Improved Precision}
\label{sec:exp}
We recently demonstrated the feasibility of Casimir pressure gradient measurements with the parallel plate \cannex{} setup~\cite{Sedmik:2018}. In the present article, we propose several improvements to the experiment that will lead to a significant increase in sensitivity and reduce the influence of systematic~effects.

\begin{figure}[!h]
\centering
\includegraphics{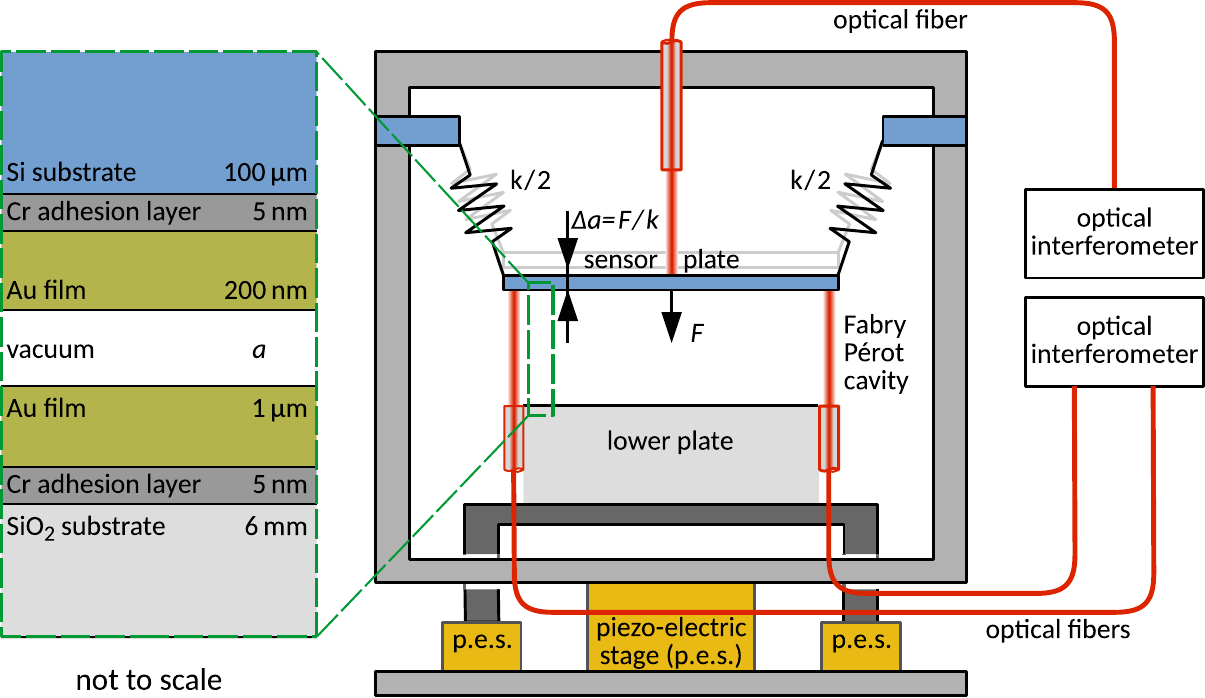}
\caption{\textls[-20]{Simplified two-dimensional schematic representation of our setup with included improvements} proposed in the main text. The material composition of the surfaces is shown separately on the left.\label{fig:setup}}
\end{figure}

In our setup shown in \figref{fig:setup}, the parallel plate geometry is implemented by a rigid vertical SiO$_2$ cylinder (lower plate) of radius $R=5.742\,$mm separated by a vacuum gap of width $a$ from a movable (upper) sensor plate that can be described as a lumped mass-spring system~\cite{Almasi:2015a}. This sensor with an interacting area $A$, elastic constant $k$, and effective mass $m_{\rm eff}$ has a free resonance frequency $\omega_0=\sqrt{k/m_{\rm eff}}$.
In~\cite{Sedmik:2018} an interferometric detection scheme was used to measure the gradient of the Casimir pressure. Here, we propose to implement the same scheme to measure both pressure gradients and the pressure acting between the two flat plates. Gradients $P^\prime(a)\equiv\partial P/\partial a$ are detected using a phase-locked loop (PLL) that senses shifts
\begin{align}
\Delta \omega=\sqrt{\omega_0^2-\frac{A}{m_{\rm eff}}\diff{P}{a}}-\omega_0
\label{eq:freq_shift}
\end{align}
of the sensor's resonance frequency---a technique widely used in the literature~\cite{Albrecht:1991,Decca:2003,Jourdan:2009,Chang:2012,Xu:2018}.
In the absence of thermal drifts, the pressure $P$ between our flat surfaces can be measured by monitoring the extension $\Delta a=PA/k$ of the spring-mass system using the same interferometer. For~the determination and maintenance of parallelism, however, we recently used a feedback mechanism based on the capacitance between the two plates. While this scheme has been demonstrated to work in principle, practice has shown that vibrations in combination with the high $Q$-factor of our sensor lead to unacceptably long integration times, and associated susceptibility to thermal drifts~\cite{Sedmik:2018}. To improve the performance and reach the full potential of the setup, we propose to replace the capacitive scheme by an interferometric one. As shown in \figref{fig:setup}, three Fabry P\'erot interferometers below the sensor plate (created by the end faces of optical fibers and the reflective surface of the sensor plate) monitor the distance $a$ at different positions around the lower plate. (Please note that in the two-dimensional scheme in the figure, only two of these cavities are depicted.) The three fiber ends can be polished together with the lower plate to exactly match the surface of the latter. In such a system, the three lower interferometers could measure the tilt and frequency shift at all times. Synchronously, the upper (fourth) interferometer, being mechanically connected to the sensor frame, can be used to measure the extension $\Delta a$ of the sensor and, hence, the pressure acting on it.

Besides this conceptual change several technical improvements are planned, which will result in the following: First, we aim to improve the seismic attenuation by a factor of 10 around $f_0=\omega_0/(2\pi)$ with respect to the present performance by means of active control techniques~\cite{Blom:2015}. This would not only reduce the direct influence on measurements via non-linear effects and rms noise, but would also improve the stability of various feedback circuits in a nontrivial way. Second, a newly designed distributed thermal control concept will guarantee mK stability throughout the setup, thereby eliminating drift and uncertainty in the sensor characteristics. Third, electrostatic patch effects that create a systematic pressure background could be reduced using in situ Ar ion bombardment~\cite{Xu:2018}. In~the following, we analyze the major sources of experimental uncertainty and show the predicted effect of the mentioned improvements.
\subsection{Sources of Error}
\label{sec:exp:errors}
Here, we consider experimental errors in the pressure and its gradient due to vibrations, the determination of displacement, and frequency shift of our sensor, variations in the temperature, tilt of the plates, and discuss the role of electrostatic patch effects.

Previously, we identified vibrations as a major source of error and gave respective tolerable limits for pressure measurements~\cite{Sedmik:2018}. Using a five-axis seismic attenuation system, we achieved a damping factor of $60$ in vertical direction around the sensor resonance $f_0=10.24\,$Hz. The residual vibrations caused at times a peak displacement noise $\delta a_{p}\approx 300\,$nm of the sensor relative to the lower plate, which proved to be a severe nuisance during the measurements. For pressure gradients, the main influence is via non-linear effects~\cite{Albrecht:1991,Antezza:2004}. For a function $\phi(a)$ representing either the pressure ($\phi=P$) or its gradient ($\phi=\partial P/\partial a$), we expand
\begin{align}
\phi(a+\delta a)\approx \phi(a)+\langle\delta a\rangle\frac{\partial}{\partial a}\phi(a)+\inv{2}\langle \delta a^2\rangle\frac{\partial^2}{\partial a^2}\phi(a)
\end{align}
 for small sensor movements $\delta a$. Recognizing that $\langle\delta a\rangle\equiv 0$, as usual for stochastic fluctuations, the main contribution to the error comes from the second order term. Thus, for the non-linear shifts in the pressure and its gradient, we therefore have
 \begin{align}
  \delta P_{\rm nl}\approx\inv{2}\langle \delta a^2\rangle\frac{\partial^2}{\partial a^2}P(a)\,,\quad \text{and}\quad\delta\left(\frac{\partial P}{\partial a}\right)_{\rm nl}\approx\inv{2}\langle \delta a^2\rangle\frac{\partial^3}{\partial a^3}P(a)\,,
 \end{align}
 respectively. Similarly, the rms pressure and pressure gradient (entering measurements mainly as a nuisance within the mechanical bandwidth of the sensor $2\delta f_{\rm BW}=2.8\,$mHz around $f_0$) are computed~from 
 the spectral sensor movement $\delta a(f)$ via
 \begin{align}
\delta \phi_{\rm rms}\approx\frac{\partial \phi(a)}{\partial a}\left[\int_{f_0-\delta f_{\rm BW}}^{f_0+\delta f_{\rm BW}}\!{\rm d}f\,\delta a^2(f)\right]^{1/2}\,.
 \end{align}

 We note that vibrational noise also hampers the convergence of various feedback circuits and thereby influences the achievable sensitivity in a nontrivial way. Experience has shown that such effects are negligible below a peak amplitude of $\sim 20\,$nm.

Another fundamental source of error is the uncertainty in various parameters involved in the evaluation of the frequency shift. We consider the following calibration procedure. At large separation $a_{cal}\approx80\,$\um{} we apply the AC electrostatic excitation voltages $V_{\rm ex}$ and $V_{\rm AC}$ driving the sensor resonance and the surface potential compensation circuit, respectively~\cite{Sedmik:2018}. At this separation, the Casimir pressure $P$ is negligibly small with respect to the electrostatic pressure $P_{\rm ES}$. The free resonance frequency
\begin{align}
 \omega_0=\left(\omega_{0P}^2 + \frac{A}{m_{\rm eff}}\left[\left.\diff{P_{\rm ES}}{a}\right|_{a_{cal}} + \left.\diff{P}{a}\right|_{a_{\rm cal}}\right]\right)^{1/2} \quad\text{with}\quad\diff{P_{\rm ES}}{a}= \frac{\vare_0}{2 a_{\rm cal}^3}\left( V_{\rm AC}^2+V_{\rm ex}^2\right)
 \end{align}
 is determined from a measurement of the resonance frequency $\omega_{0P}$ under the influence of the well-known electrostatic and Casimir forces. Here, $\vare_0$ is the dielectric permittivity of vacuum. The effective mass $m_{\rm eff}$ is determined during a separate sweep recording $\omega_{0P}$ as a function of an applied DC electrostatic potential. Eventually, we evaluate the resonance frequency shift measured at different separations to obtain the Casimir pressure gradient from \eqnref{eq:freq_shift} after subtraction of the electrostatic pressure gradient $\partial P_{\rm ES}/\partial a$. For pressure measurements at the same separations, we~evaluate the detected extension $\Delta a$ of the sensor that is related to the total pressure $P_{\rm tot}=P+P_{\rm ES}=\Delta a\, \omega_0^2m/A$. The achievable sensitivity for both types of measurement is limited by the uncertainties in all experimental quantities entering the respective evaluation. Numerical values for these and some other uncertainties considered below are given in \tabref{tab:exp_errors}.

Variations in the temperature contribute in two ways to the experimental error. First, different material expansion coefficients influence the separation between the two interacting surfaces by roughly $63\,$nm/K---an effect most influential at smaller $a$. Second, the temperature influences the Youngs modulus of our sensor, which leads to an additional error in $a$, but also offsets the resonance frequency and, thereby, mimics a pressure gradient.

We also consider errors due to tilt of the plates with respect to each other. For small angles $\delta\alpha$ between the plates, we may estimate the influence on the pressure and its gradient by averaging
\begin{align}
\delta \phi(a,\delta\alpha)\approx\int_A \!{\rm d}A\,\phi[a(x,y,\delta\alpha)]\,,
\end{align}
over the sensor area $A$ with the local separation $a(x,y,\delta\alpha)$ deviating from its nominal value $a$ due to the tilt. Numerical calculations show that the respective relative corrections to the pressure and its gradient are both of the order $(\delta\a R/a)^2$, in agreement with the literature~\cite{Bordag:2014}. For the achievable values of $\delta a$ given in \tabref{tab:exp_errors}, these corrections are negligible.

As has been discussed in Ref.~\cite{Sedmik:2018}, electrostatic effects can have a significant influence. While we recently compared our measurements with the model~\cite{Speake:2003}, we now use the model ~\cite{Behunin:2012} that has been shown to describe the observed forces more realistically. The average patch size $\langle\ell\rangle=0.82\,$\um{} and the value for $\langle\ell^2\rangle$ mentioned in \tabref{tab:exp_errors} were derived from the auto-correlation of actual Kelvin probe data for our surfaces. As $\langle\ell\rangle$ is much smaller than the plate separation, we can use the approximation \begin{align}
P_{patch}(a)\approx \frac{3\zeta(3)\vare_0V_{\rm rms}^2\langle\ell^2\rangle}{4a^4}\,.
\end{align}

 Here, $V_{\rm rms}=0.634\,$mV, and $\zeta(z)$ is the Riemann zeta function. Please note that $P_{\rm patch}$ represents a systematic effect that can be characterized and removed from experimental data.
\begin{table}[!h]
 \centering
 \caption{Values for the experimental parameters and uncertainties for the setup in its present configuration and after the proposed improvements.\label{tab:exp_errors}}
 \begin{tabular}{lccc}
  \toprule
  \textbf{Parameter} & {\textbf{Present Value}} &{\textbf{Improved Value}}\\
  \midrule
  Separation uncertainty $\delta a_{\rm cal}$ &2.0 &0.5&nm\\
  Mass calibration error $\delta m_{\rm eff}$& 1.0 & 0.01&\%\\
  Area uncertainty $\delta A$&{$5\times10^{-4}$}&{$5\times10^{-4}$}&cm$^2$\\
  Applied AC voltages uncertainty& 10&1&\SI{}{\micro\volt}\\
  Frequency measurement error &$10^{-5}$&{$4\times10^{-7}$}&Hz\\
  Vibration amplitude at $f_0$ &300&20&nm\\
  Patches $V_{\rm rms}$ & 1.28 & 0.64&V\\
  Patches $\langle \ell^2\rangle$ &7.3 &2.4&\um{}$^2$\\
  Tilt angle $\delta \a$&3.0 & 0.1&\SI{}{\micro\radian}\\
  Temperature stability &3.0 & 0.5&mK\\
  \bottomrule
 \end{tabular}
\end{table}
\subsection{Sensitivity Estimation}
\label{sec:exp:sens_est}
\begin{figure}[!h]
\centering
 \includegraphics[scale=0.9]{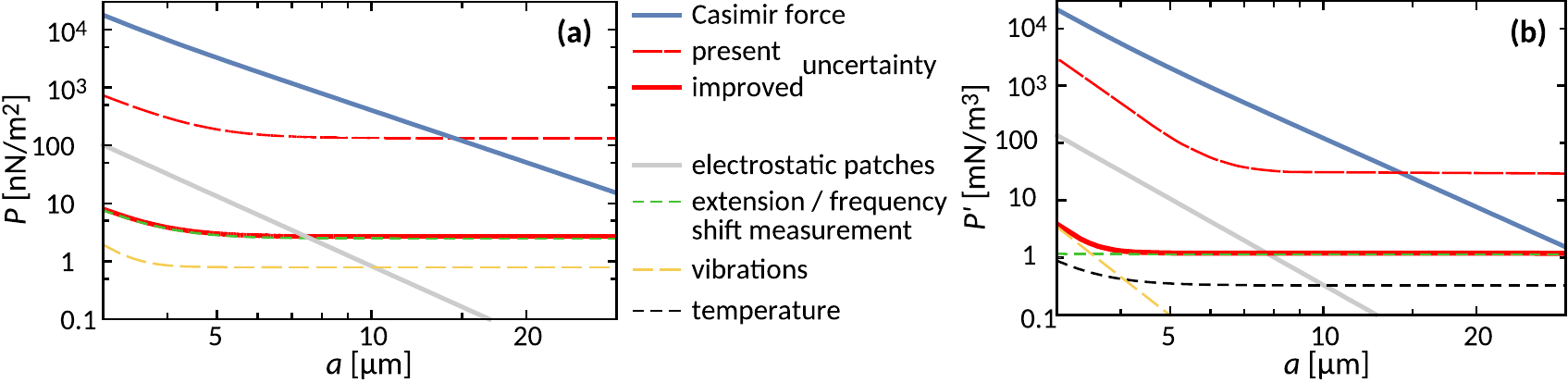}
 \caption{Estimated experimental uncertainties for the present and improved versions of the setup as a function of separation between the two parallel plates. (\textbf{a}) Pressure measurements using the upper interferometer. (\textbf{b}) Pressure gradient measurements using the lower interferometer.\label{fig:uncertainty}}
\end{figure}

Based on the models described in \secref{sec:exp:errors}, we have calculated the expected level of experimental uncertainty in measurements of the pressure and its gradient for both the present situation and presuming a successful implementation of the proposed improvements. For these calculations, we have assumed the parameters listed in \tabref{tab:exp_errors}. As can be seen in \figref{fig:uncertainty}a, the largest uncertainty in force measurements comes from the determination of the sensor extension, which includes contributions from the calibration and interferometry. Vibrations play, under the assumption that the proposed measures work as expected, a minor role. For force gradients (\figref{fig:uncertainty}b) the determination of the frequency shift is the limiting factor at separations larger than $\sim\SI{4}{\micro\meter}$, while at smaller separations sensitivity is limited by vibrations. Temperature variations are influential on force gradient measurements for all separations, as they modify the Youngs modulus of our sensor. For comparison, we also plot the achievable uncertainty for the present version of the setup (dashed upper red lines), which are determined by the same factors as for the improved version.
Please note that the patch pressure can be characterized separately and removed from the data.
The results for the improved uncertainties in \figref{fig:uncertainty} are based on very conservative estimates. In the experiment, especially at larger separations, the uncertainty in the pressure could be reduced statistically by longer measurements. For further calculations in \secref{sec:constraints}, we therefore assume a pressure sensitivity of \SI{1}{\nano\newton/\meter^2} (corresponding to \SI{0.1}{\pico\newton}) at $a\geq\SI{4}{\micro\meter}$, which is slightly higher than the value corresponding to the bottom red line in \figref{fig:uncertainty}a. The uncertainty in the pressure gradient is mainly determined by the resolution of the PLL's frequency measurement. Here, the possibility for a statistical reduction of the uncertainty may be more problematic and therefore we use the pressure gradient sensitivity given by the bottom red line in \figref{fig:uncertainty}b for calculations in \secref{sec:constraints}.
\section{Possibilities to Measure Thermal Effects in the Casimir Force}
\label{sec:thermal}
As mentioned in \secref{sec:intro}, most of the already performed experiments on measuring the Casimir interaction
exploited the sphere-plate geometry. In doing so, the measured quantity was either the Casimir force acting between
a sphere and a plate (in the static measurement scheme) or its gradient (in the dynamic measurement scheme).
Due to the proximity force approximation, the latter quantity can be recalculated into the effective Casimir pressure
between two parallel plates \cite{Bordag:2014,Klimchitskaya:2009a}.
There is only one modern experiment on the direct measurement of the
Casimir pressure between two parallel plates \cite{Bressi:2002}, but it is not of sufficient precision to observe the
thermal effects (an attempt to measure the Casimir effect between two
parallel Al-coated plates at separations larger than a few micrometers
was unsuccessful due to the presence of large background forces \cite{42a}).

The distinctive feature of the \cannex{} test of the quantum vacuum is that it can be adapted for
simultaneous measurements of the Casimir pressure between two parallel plates and its gradient
(see \secref{sec:exp}).

The Lifshitz formula for the Casimir pressure between two material plates spaced at a separation $a$ at temperature $T$ is given
by \cite{Bordag:2014,Klimchitskaya:2009a}
\begin{equation}
{P}(a)=
-\frac{k_BT}{\pi}\sum_{l=0}^{\infty}
\vphantom{\sum}^{'}\int_0^{\infty}k_{\bot}dk_{\bot}q_l
 \sum_{\alpha}
\frac{r^2_{\alpha}(i{\xi}_l, k_{\bot})}{e^{2aq_l}-r^2_{\alpha}(i{\xi}_l, k_{\bot})}.
\label{eq0}
\end{equation}

Here, it is assumed that the parallel plates made of a nonmagnetic material described by the dielectric permittivity
$\varepsilon(\omega)$ are in thermal equilibrium with the environment at temperature $T$ and the following
notations are introduced.
The Boltzmann constant is $k_B$, the prime on the summation sign  divides the term with $l=0$ by $2$,
$k_{\bot}$ is
the magnitude of the projection of the wave vector on the plane of the plates,
 $\xi_l=2{\pi}k_BTl/\hbar$ with $l=0,\,1,\,2,\,\ldots\,$ are the Matsubara frequencies , and
$q_l=(k_{\bot}^2+\xi_l^2/c^2)^{1/2}$.

The reflection coefficients $r_{\alpha}$  are defined for two independent polarizations of the electromagnetic
field, transverse magnetic ($\alpha={\rm TM}$) and transverse electric
($\alpha={\rm TE}$). Explicitly they are given by
\begin{equation}
r_{\rm TM}(i\xi_l,k_{\bot})=\frac{\varepsilon_lq_l-k_l}{\varepsilon_lq_l+k_l},
\qquad
r_{\rm TE}(i\xi_l,k_{\bot})=\frac{q_l-k_l}{q_l+k_l},
\label{eq2}
\end{equation}
\noindent
where
\begin{equation}
k_l\equiv k(i\xi_l,k_{\bot})=\left(k_{\bot}^2+\varepsilon_l\frac{\xi_l^2}{c^2}\right)^{1/2}
{\!\!\!\!},
\qquad
\varepsilon_l\equiv\varepsilon(i\xi_l).
\label{eq3}
\end{equation}

 By differentiating \eqnref{eq0} with respect to separation between
the plates $a$, one obtains the gradient of the Casimir pressure
\begin{equation}
{P}^{\prime}(a)=
\frac{2k_BT}{\pi}\sum_{l=0}^{\infty}
\vphantom{\sum}^{'}\int_0^{\infty}k_{\bot}dk_{\bot}q_l^2
 \sum_{\alpha}
\frac{r^2_{\alpha}(i{\xi}_l, k_{\bot})e^{2aq_l}}{\left[e^{2aq_l}-r^2_{\alpha}(i{\xi}_l, k_{\bot})\right]^2}.
\label{eq1}
\end{equation}
\noindent

Equations (\ref{eq0}) and (\ref{eq1}) take an exact account of the effects of
finite conductivity of the plate metal. As to corrections due to surface
roughness, at separations exceeding $3~\upmu$m they are much smaller than
an error in the pressure measurements
\cite{Bordag:2014}.

For numerical computations it is convenient to introduce the dimensionless variables
\begin{equation}
y=2aq_l,\qquad \zeta_l=\frac{2a\xi_l}{c}.
\label{eq4}
\end{equation}

In terms of these variables Equations (\ref{eq0}) and (\ref{eq1}) take the form
\begin{equation}
{P}(a)=-\frac{k_BT}{8\pi a^3}\sum_{l=0}^{\infty}
\vphantom{\sum}^{'}\int_{\zeta_l}^{\infty}y^2dy
 \sum_{\alpha}
\frac{r^2_{\alpha}(i{\zeta}_l, y)e^{-y}}{1-r^2_{\alpha}(i{\zeta}_l, y)e^{-y}}\,,
\label{eq5a}
\end{equation}
\noindent
and
\begin{equation}
{P}^{\prime}(a)=\frac{k_BT}{8\pi a^4}\sum_{l=0}^{\infty}
\vphantom{\sum}^{'}\int_{\zeta_l}^{\infty}y^3dy
 \sum_{\alpha}
\frac{r^2_{\alpha}(i{\zeta}_l, y)e^{-y}}{\left[1-r^2_{\alpha}(i{\zeta}_l, y)e^{-y}\right]^2},
\label{eq5}
\end{equation}
\noindent
respectively, and the reflection coefficients from \eqnref{eq2} take the form
\begin{equation}
r_{\rm TM}(i\zeta_l,y)=\frac{\varepsilon_l y-
\sqrt{y^2+(\varepsilon_l-1)\zeta_l^2}}{\varepsilon_l y+\sqrt{y^2+(\varepsilon_l-1)\zeta_l^2}},
\qquad
r_{\rm TE}(i\zeta_l,y)=\frac{ y-\sqrt{y^2+(\varepsilon_l-1)\zeta_l^2}}{ y+\sqrt{y^2+(\varepsilon_l-1)\zeta_l^2}}.
\label{eq6}
\end{equation}

For application to the experimental setup of \cannex{} described in \secref{sec:exp},
computations should be made for Au plates. The dielectric permittivity of Au along the imaginary frequency
axis is obtained by means of the Kramers-Kronig relations using the available tabulated optical data for the
complex index of refraction extrapolated down to zero frequency \cite{Bordag:2014,Klimchitskaya:2009a}. According to \secref{sec:intro}, this
extrapolation can be made either by means of the Drude model taking into account the relaxation properties
of conduction electrons or the plasma model disregarding these relaxation properties. In~dimensionless
variables, the dielectric permittivity of the Drude model along the imaginary frequency axis is given by
\begin{equation}
\varepsilon_l^{(D)}=1+\frac{{\tilde\omega_p}^2}{\zeta_l(\zeta_l+\tilde\gamma)}\, ,
\label{eq7}
\end{equation}
\noindent
where the dimensionless plasma frequency $\tilde\omega_p$ and relaxation parameter $\tilde\gamma$ are
connected with the dimensional ones by   $\tilde\omega_p=2a\omega_p/c$ and $\tilde\gamma=2a\gamma/c$.
For Au the standard values $\hbar\omega_p=9.0~$eV and $\hbar\gamma=35~$meV are used here.
The dielectric permittivity of the plasma model $\varepsilon_l^{(p)}$ is obtained from \eqnref{eq7} by putting
$\tilde\gamma=0$. Please note that for computations at separations $a\geq 3~\upmu$m performed below
the optical data contribute negligibly small, so that the obtained gradients of the Casimir pressure
are mostly determined by the extrapolations to lower frequencies.

Before presenting the computational results, we note that in the high-temperature limit
$T\gg T_{\rm cr}=\hbar c/(2k_B a)$ all the terms in Equations (\ref{eq5a}) and (\ref{eq5}) with $l\geq 1$ are
exponentially small and both the Casimir pressure and its gradient are given predominantly by the terms with $l=0$.
The zero-frequency term of the Lifshitz formula takes different forms depending on the extrapolation used.
If the plasma model  $\varepsilon_l^{(p)}$ is used for extrapolation, one obtains
\begin{align}
P_{(p)}(a)&=-\frac{k_B T}{16\pi a^3}\left[\int_0^{\infty}\!\!dy\frac{y^2e^{-y}}{1-e^{-y}}
+\int_0^{\infty}\!\!dy\frac{y^2r_{\rm TE}^2(0,y)e^{-y}}{1-r_{\rm TE}^2(0,y)e^{-y}}\right]\,
\label{eq8a}
\end{align}
\mbox{}\vspace{-6pt}and\vspace{-1pt}\mbox{}
\begin{align}
P_{(p)}^{\prime}(a)&=\frac{k_B T}{16\pi a^4}\left[\int_0^{\infty}\!\!dy\frac{y^3e^{-y}}{(1-e^{-y})^2}
+\int_0^{\infty}\!\!dy\frac{y^3r_{\rm TE}^2(0,y)e^{-y}}{[1-r_{\rm TE}^2(0,y)e^{-y}]^2}\right]\, ,
\label{eq8}
\end{align}
\noindent
where in accordance to \eqnref{eq6}
\begin{equation}
r_{\rm TE}(0,y)=\frac{y-\sqrt{y^2+{\tilde\omega_p}^2}}{y+\sqrt{y^2+{\tilde\omega_p}^2}}.
\label{eq9}
\end{equation}

Calculating the first integrals on the right-hand side of Equations (\ref{eq8a}) and (\ref{eq8})
and expanding the second ones in the powers
of a small parameter $1/\tilde\omega_p$, we find \cite{Bordag:2014}
\begin{equation}
P_{(p)}(a)=-\frac{k_B T\zeta(3)}{4\pi a^3}\left(1-\frac{6}{\tilde\omega_p}+
\frac{48}{{\tilde\omega_p}^2}\right) ,
\quad
P_{(p)}^{\prime}(a)=\frac{3k_B T\zeta(3)}{4\pi a^4}\left(1-\frac{8}{\tilde\omega_p}+
\frac{80}{{\tilde\omega_p}^2}\right)\,.
\label{eq10}
\end{equation}

If, however, the Drude model  $\varepsilon_l^{(D)}$ is used for extrapolation, one arrives at the result
\begin{equation}
P_{(D)}(a)=-\frac{k_B T}{16\pi a^3}\int_0^{\infty}\!\!dy\frac{y^2e^{-y}}{1-e^{-y}}
=-\frac{k_B T\zeta(3)}{8\pi a^3}\,
\label{eq11a}
\end{equation}
\noindent
\mbox{}\vspace{-6pt}and\vspace{-1pt}\mbox{}
\begin{equation}
P_{(D)}^{\prime}(a)=\frac{k_B T}{16\pi a^4}\int_0^{\infty}\!\!dy\frac{y^3e^{-y}}{(1-e^{-y})^2}
=\frac{3k_B T\zeta(3)}{8\pi a^4}.
\label{eq11}
\end{equation}

Please note that at $T=300\,$K \eqnsref{eq10}{eq11} are applicable under a condition
\begin{equation}
a\gg\frac{\hbar c}{2k_B T}\approx 4~\upmu\mbox{m}.
\label{eq12}
\end{equation}
\begin{figure}[!b]
\centering
\includegraphics[scale=0.92]{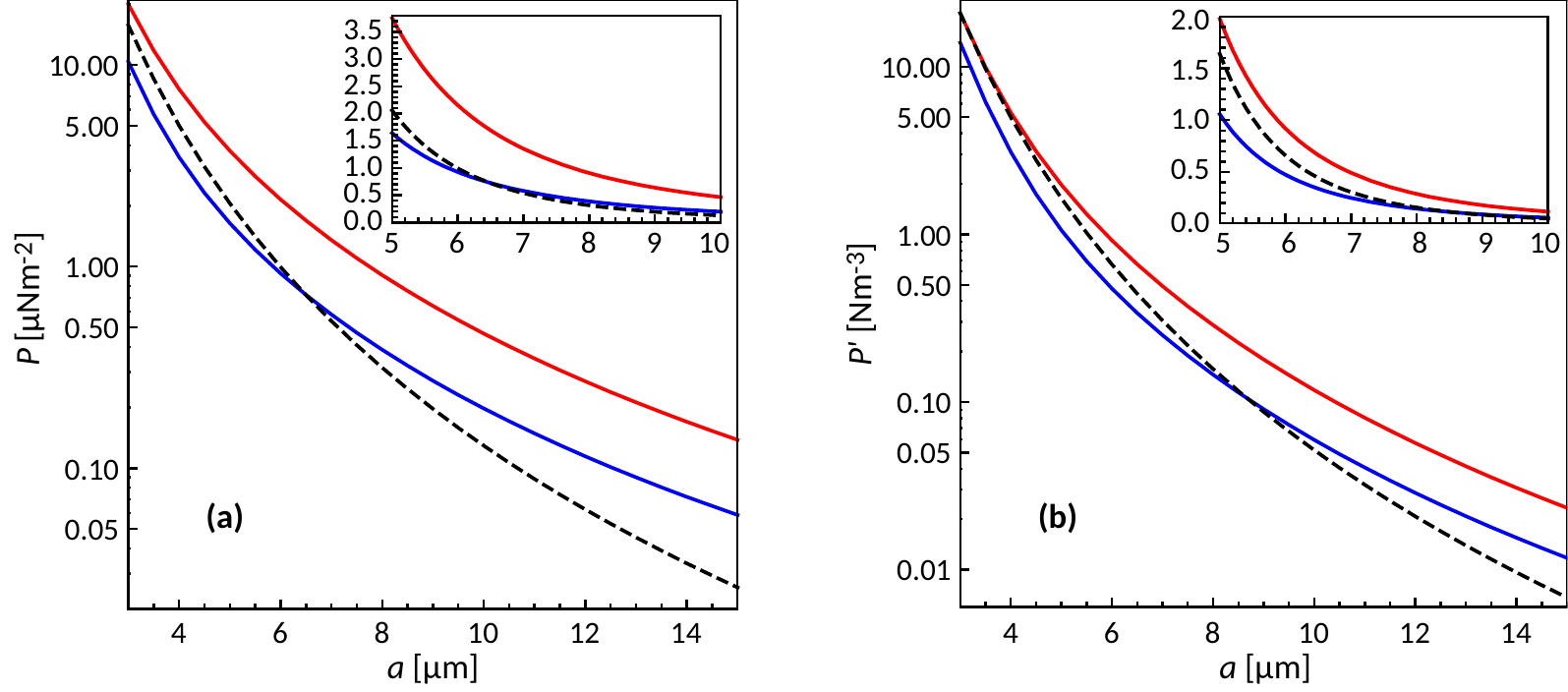}
\caption{\label{fgP1}(\textbf{a}) The magnitude of the Casimir pressure between two parallel Au plates and (\textbf{b}) the pressure
gradient are shown as  functions
of separation by the red and blue solid lines computed at $T=300~$K using the plasma and Drude extrapolations
of the optical data, respectively. The same quantities computed at $T=0$ are shown by the dashed lines. The region
of separations from 5 to 10~$\upmu$m is shown in the insets using the homogeneous scale.}
\end{figure}

Taking into account, however, that the powers in all exponentially small terms with $l\geq 1$ depend on
$-2\pi T/T_{\rm cr}$, one can see that  at $T=300~$K \eqnsref{eq10}{eq11}  lead to rather
precise results at all separations exceeding 6 or \SI{7}{\micro\meter}.

The computational results for the magnitude of the Casimir pressure and its gradient as functions of
separation between the plates are presented in \figref{fgP1}

by the red and blue solid lines computed at $T=300~$K using extrapolations of the
optical data of Au by means of the plasma and Drude models, respectively. For comparison purposes,
the computational results at $T=0$ are presented by the dashed lines. They are obtained by Equations (\ref{eq5a})
and (\ref{eq5}) where summation over the discrete Matsubara frequencies is replaced with a continuous integration.
It is taken into account that with decreasing temperature the relaxation parameter $\gamma$ quickly decreases
towards a very small residual value at $T=0$ that is determined by the defects of the crystal lattice.
As a result, the values  of $P(a)$ and  $P^{\prime}(a)$ computed at $T=0$ using the plasma and Drude models
coincide at high accuracy.

For a better visualization, in the insets to Figure~\ref{fgP1}a,b the computational results in the separation region from
5 to \SI{10}{\micro\meter} are presented using a uniform scale on the vertical axis. It is clearly seen that the
theoretical predictions from using the plasma and Drude model extrapolations of the optical data can be discriminated if to take into account the errors in measuring $P(a)$ and $P^{\prime}(a)$
in the improved version of \cannex{} discussed in \secref{sec:exp}. Please note that if the Drude model extrapolation is used the thermal
  effect in the  Casimir pressure and its gradient vanishes at approximately 6.4 and $\SI{8.7}{\micro\meter}$ separations,
respectively, where the blue and dashed lines intersect.

\begin{figure}[!hb]
\centering
\includegraphics[scale=0.932]{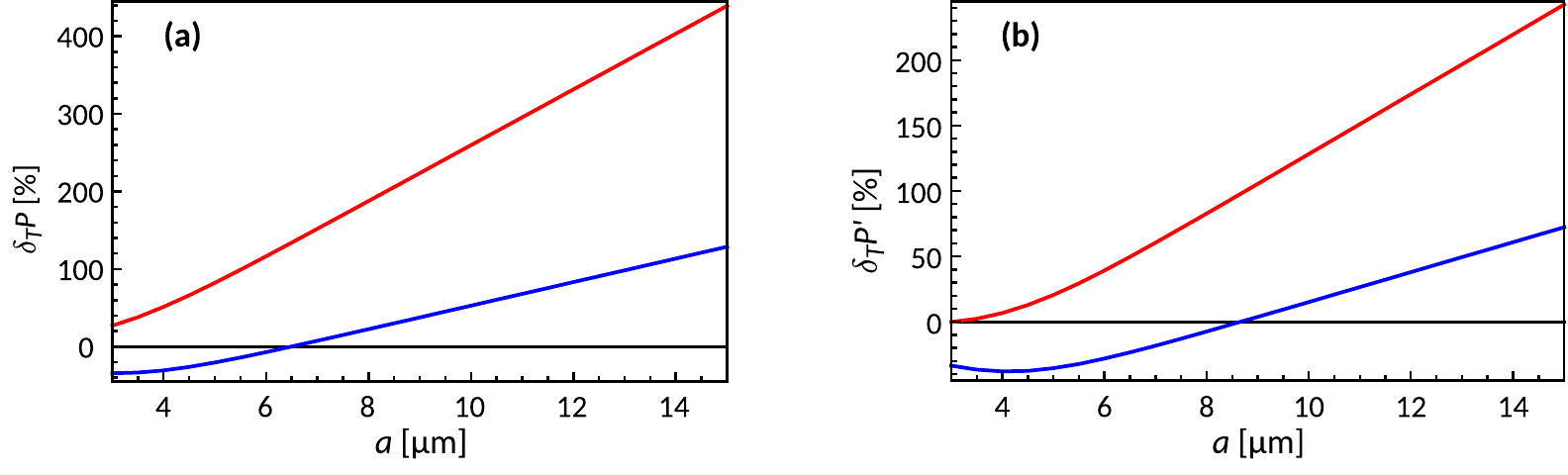}
\caption{\label{fgP2}The relative thermal corrections to (\textbf{a}) the Casimir pressure between two parallel Au plates
and (\textbf{b}) its gradient is shown  as functions of separation by the red and blue  lines computed at $T=300~$K using the plasma and Drude extrapolations
of the optical data, respectively.}
\end{figure}

To determine the specific role of thermal effects in the Casimir pressure and its gradient, we have also computed
the relative thermal corrections defined as
\begin{equation}
\delta_T P(a)=\frac{ P(a,T)- P(a,0)}{ P(a,0)},
\qquad
\delta_T P^{\prime}(a)=\frac{ P^{\prime}(a,T)- P^{\prime}(a,0)}{ P^{\prime}(a,0)}\,,
\label{eq13}
\end{equation}
where on the right-hand sides we indicated the dependence on temperature explicitly.
\noindent
Computational results for $\delta_T P$ and $\delta_T P^\prime$ are presented in Figure~\ref{fgP2}a,b
 as functions of separation by the red and blue lines for the cases
when extrapolation of the optical data for Au to lower frequencies is made by means of the plasma and Drude models,
respectively. As is seen in these figures, within the separation range from 3 to $15~\upmu$m
 the thermal effects make a considerable contribution to the Casimir pressure and its gradient.
If the plasma model extrapolation is used,
it reaches 440\% and 240\% of the zero-temperature pressure and its gradient, respectively, at $a=15~\upmu$m.
When using the extrapolation
by means of the Drude model, the relative thermal effect in the Casimir pressure
 varies from approximately $-$34\% at  $a=3~\upmu$m to
130\% at  $a=15~\upmu$m and in the pressure gradient
  from approximately $-$40\% at  $a=4~\upmu$m to 60\% at  $a=15~\upmu$m.
 In this case, the  thermal effect in the Casimir pressure vanishes  at $a\approx 6.4~\upmu$m
and in its gradient at $a\approx 8.7~\upmu$m in agreement with
Figure~\ref{fgP1}a,b, respectively .

\section{Prospective Constraints on Non-Newtonian Gravity and Axion-Like Particles}
\label{sec:constraints}
As mentioned in \secref{sec:intro}, the Casimir forces originating from the quantum vacuum form a background for
testing the Yukawa-type corrections to Newton's gravitational law and for searching the axion-like particles.
It is convenient to parametrize the Yukawa-type potential between two point masses $m_1$ and $m_2$
situated at the points $\mbox{\boldmath$r$}_1$ and $\mbox{\boldmath$r$}_2$ as \cite{Fischbach:1999}\\*[-6pt]\mbox{}
\begin{equation}
V_{\rm Yu}(r)\equiv V_{\rm Yu}(|\mbox{\boldmath$r$}_1-\mbox{\boldmath$r$}_2|)=-\alpha
\frac{Gm_1m_2}{|\mbox{\boldmath$r$}_1-\mbox{\boldmath$r$}_2|}\exp\!\left(
-\frac{(|\mbox{\boldmath$r$}_1-\mbox{\boldmath$r$}_2|}{\lambda}\right).
\label{eq14}
\end{equation}

Here, $\alpha$ and $\lambda$ are the interaction constant  and the range of Yukawa interaction, and
$G=6.674\times 10^{-11}\,\mbox{m}^3\,\mbox{kg}^{-1}\,\mbox{s}^{-2}$ is the Newtonian gravitational constant
 {(we }note that in the experimental configuration under consideration one can
neglect by the Newtonian gravitational pressure because it is less than
an error in measurements of the Casimir pressure).

The Yukawa-type pressure between the top and bottom plates in the experimental setup of \secref{sec:exp} should be
calculated taking into account the layer structure of both plates shown in \figref{fig:setup}. The top plate is made of high-resistivity Si of density
$\rho_{\rm Si}=2.33~\mbox{g\,cm}^{-3}$
coated with a layer of Cr of density
$\rho_{\rm Cr}=7.15~\mbox{g\,cm}^{-3}$
and thickness $d_{\rm Cr}=5~$nm followed by a layer of Au of density
$\rho_{\rm Au}=19.3~\mbox{g\,cm}^{-3}$
and thickness $d_{\rm Au}^t=\SI{200}{nm}$.
The thickness of Si substrate ($100~\upmu$m) is sufficiently large to treat it as a semispace.
The bottom plate is made of  SiO$_2$ quartz crystal with the density
$\rho_{q}=2.64\,\mbox{g\,cm}^{-3}$
coated with a layer of Cr of  thickness $d_{\rm Cr}=5\,$nm followed by a layer of Au of
thickness $d_{\rm Au}^b=\um{1}$.
The thickness of SiO$_2$ substrate ($6\,$mm) again allows to consider it as a semispace.

Now we assume that one mass $\rho dV_1$ belongs to the top plate and another one  $\rho dV_2$
to the bottom one and integrate \eqnref{eq14} over the volumes of both parallel plates separated by a distance $a$
taking into account their layer structure. Calculating the negative derivative of the obtained interacting energy with
respect to $a$, we find the Yukawa force and finally the pressure \cite{Decca:2005a}\\*[-3pt]\mbox{}
\begin{equation}
P_{\rm Yu}(a)=-2\pi G\alpha\lambda^2e^{-a/\lambda}\Phi(\lambda),
\label{eq15}
\end{equation}
\noindent
where the function $\Phi$ is defined as\\[-15pt]\mbox{}
\begin{align}
\Phi(\lambda)=&\left[\rho_{\rm Au}-(\rho_{\rm Au}-\rho_{\rm Cr})\exp\,\left(-\frac{d_{\rm Au}^t}{\lambda}\right)
-(\rho_{\rm Cr}-\rho_{\rm Si})\exp\,\left(-\frac{d_{\rm Au}^t+d_{\rm Cr}}{\lambda}\right)\right]
\nonumber \\[-1pt]
&\times
\left[\rho_{\rm Au}-(\rho_{\rm Au}-\rho_{\rm Cr})\exp\,\left(-\frac{d_{\rm Au}^b}{\lambda}\right)
-(\rho_{\rm Cr}-\rho_{q})\exp\,\left(-\frac{d_{\rm Au}^b+d_{\rm Cr}}{\lambda}\right)\right].
\label{eq16}
\end{align}

The gradient of the Yukawa pressure is obtained by differentiating \eqnref{eq15} with respect to $a$
\begin{equation}
P_{\rm Yu}^{\prime}(a)=2\pi G\alpha\lambda e^{-a/\lambda}\Phi(\lambda).
\label{eq17}
\end{equation}

Now the constraints on the parameters $\alpha$ and $\lambda$ of Yukawa-type interactions can be obtained from the inequalities
\begin{equation}
\left| P_{\rm Yu}(a)\right|<\Delta P(a), \qquad
P_{\rm Yu}^{\prime}(a)<\Delta P^{\prime}(a),
\label{eq18}
\end{equation}
\noindent
where $\Delta P$ and $\Delta P^{\prime}$ are the total experimental errors in the measured Casimir pressure and its
gradient estimated in \secref{sec:exp}. The meaning of \eqnref{eq18} is that the experimental data are found in agreement with
theoretical predictions for the Casimir pressure and its gradient and no extra contribution of unknown origin was observed.

\begin{figure}[!ht]
\centering
\includegraphics[scale=0.93]{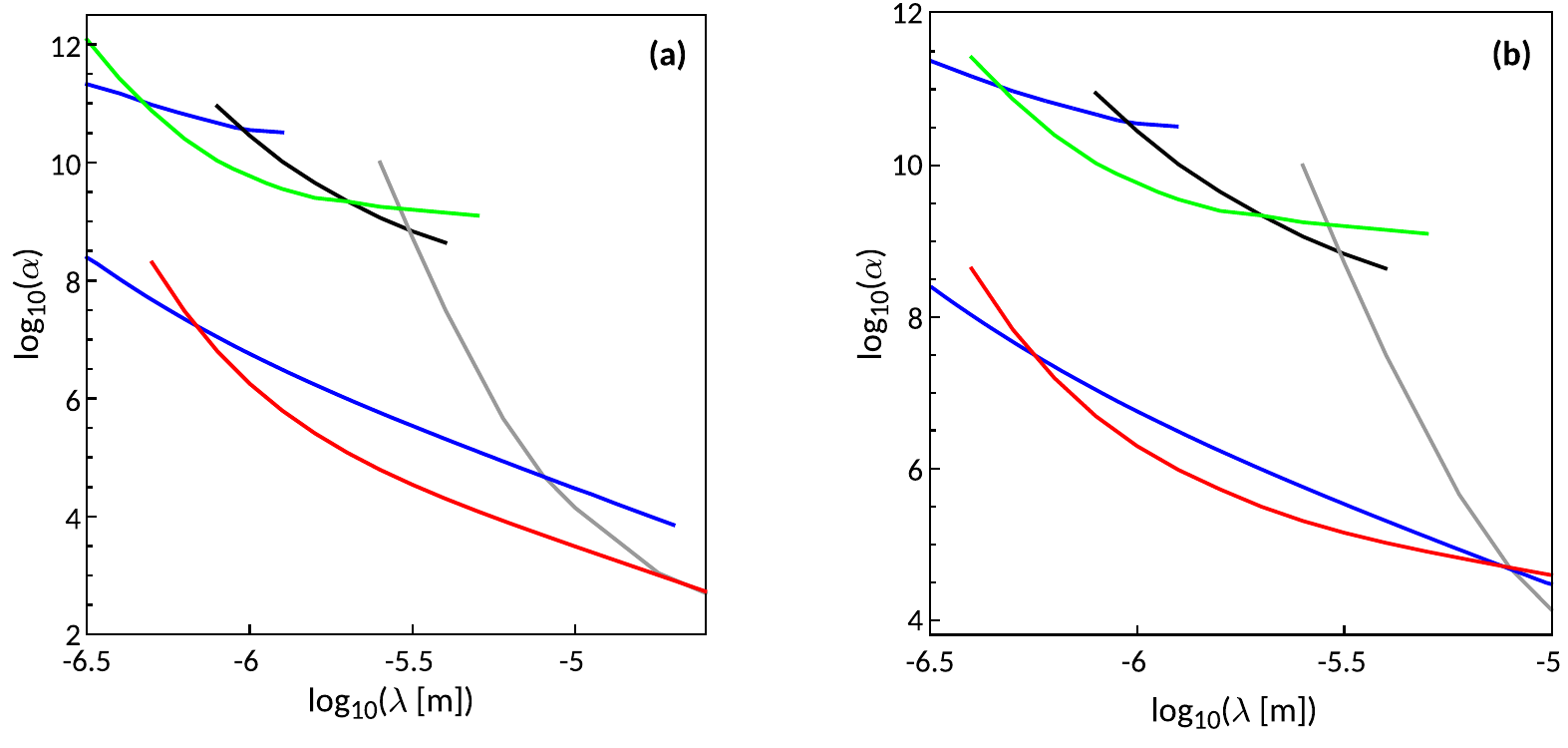}
\caption{\label{fgC1}Constraints on the parameters of Yukawa-type interaction obtained from two Casimir-less
experiments, from measuring the difference in lateral forces, from the torsion pendulum experiment and from the
Cavendish-type experiments are shown by the top and bottom blue lines, green, black and gray lines,
respectively. The proposed constraints which could be obtained from the \cannex{} test of the quantum vacuum
when measuring (\textbf{a}) the Casimir pressure and (\textbf{b}) its gradient are indicated by the red lines.}
\end{figure}

To estimate the strength of prospective constraints, which can be obtained from the improved version of \cannex{},
we use the thickness of the top and bottom Au layers $d_{\rm Au}^t=200~$nm and  $d_{\rm Au}^b=1~\upmu$m
and the total experimental errors $\Delta P(a)=\SI{1}{\nano\newton/\meter^2}$ and $\Delta P^{\prime}(a)$ corresponding to the bottom red line in~\figref{fig:uncertainty}b. The computational results for $\alpha$ and $\lambda$ obtained by using the first and second
inequalities in \eqnref{eq18} are shown by the red lines in Figure~\ref{fgC1}a,b, respectively.
In so doing, the regions of $(\alpha,\lambda)$-planes above each line are excluded and below each line are allowed.

For comparison purposes, we present in  Figure~\ref{fgC1}a,b the strongest constraints
obtained in the same interaction range from other experiments. The top and bottom blue lines demonstrate
the constraints found from the results of the Casimir-less experiment \cite{Decca:2005} and its improved
version \cite{Chen:2016}, respectively. Both experiments were performed by means of a micromechanical
torsional oscillator. The green line shows the constraints obtained from measuring the difference in
lateral forces \cite{Wang:2016}. The constraints indicated by the black line are found in \cite{Masuda:2009} from the torsion
pendulum experiment. Finally, the constraints of the gray line follow from the Cavendish-type experiments
performed at short separations \cite{Chiaverini:2003,Smullin:2005,Geraci:2008}.

As is seen  in  Figure~\ref{fgC1}a,b, the strongest constraints obtained up to date
follow from the improved Casimir-less experiment \cite{Chen:2016} (the bottom blue line). It is seen also
that the largest strengthening of these constraints, which could be reached from the \cannex{} test of
quantum vacuum, follows from measurements of the Casimir pressure (this is because the pressure gradient
is linear in $\lambda$ whereas the pressure is quadratic in $\lambda$).  The prospective constraints are
stronger by up to a factor of 10 over a wide interaction range.

Next we consider the prospective constraints on the axion-to-nucleon coupling constants which could be
obtained from the \cannex{} test of the quantum vacuum. Taking into account that the exchange of one axion between
two nucleons results in the spin-dependent interaction potential \cite{Adelberger:2003a} and the test bodies in this
experiment are not polarized, any additional  force of the axion origin could arise due to two-axion exchange.

Below we deal with axion-like particles coupled to nucleons by means of pseudo-scalar interaction
Lagrangian \cite{Kim:1987}. In this case, the effective potential between two nucleons,  spaced
at the points $\mbox{\boldmath$r$}_1$ and   $\mbox{\boldmath$r$}_2$ of the top and bottom plates,
arising due to exchange of two axions, takes the form \cite{Adelberger:2003a,Ferrer:1999}
\begin{equation}
V_{an}(r)\equiv V_{an}(|\mbox{\boldmath$r$}_1-\mbox{\boldmath$r$}_2|)=-
\frac{g_{an}^4}{32\pi^3}\frac{\hbar^2 m_a}{m^2}
(\mbox{\boldmath$r$}_1-\mbox{\boldmath$r$}_2)^{-2}{ K}_1\!\left(
\frac{2m_ac|\mbox{\boldmath$r$}_1-\mbox{\boldmath$r$}_2|}{\hbar}\right).
\label{eq19}
\end{equation}

Here, $g_{an}$ is the dimensionless coupling constant of an axion to a nucleon (we assume that the
coupling constants to a neutron and a proton are equal \cite{Adelberger:2003a}), the mean of the proton and neutron masses is
denoted as $m$, the axion mass is $m_a$, ${ K}_1(z)$ is the modified Bessel function of the second kind
and it is assumed that $r\gg\hbar/(mc)$.

Similar to the case of a Yukawa-type potential, the additional pressure between the test bodies due to
two-axion exchange, is obtained by integrating \eqnref{eq19} over the volumes of both plates with account of their
layer structure, calculate the negative derivative of the obtained result with respect to $a$ and finally find the pressure
(see \cite{Bezerra:2014b} for details)
\begin{equation}
P_{an}(a)=-\frac{\hbar^3}{2m^2m_{\rm H}^2c}\int_{1}^{\infty}du\frac{\sqrt{u^2-1}}{u^2}
\Psi\left(\frac{m_acu}{\hbar}\right)
\exp\,\left(-\frac{2m_acau}{\hbar}\right).
\label{eq20}
\end{equation}

Here, $m_{\rm H}$ is the mass of atomic hydrogen and the function $\Psi(x)$ is defined as
\begin{eqnarray}
&&
\Psi(x)=\left[C_{\rm Au}(1-e^{-2xd_{\rm Au}^t})+C_{\rm Cr}e^{-2xd_{\rm Au}^t}(1-e^{-2xd_{\rm Cr}})
+C_{\rm Si}e^{-2x(d_{\rm Au}^t+d_{\rm Cr})}\right]
\nonumber \\
&&~~\times
\left[C_{\rm Au}(1-e^{-2xd_{\rm Au}^b})+C_{\rm Cr}e^{-2xd_{\rm Au}^b}(1-e^{-2xd_{\rm Cr}})
+C_{q}e^{-2x(d_{\rm Au}^b+d_{\rm Cr})}\right],
\label{eq21}
\end{eqnarray}
\noindent
where the coefficients $C$ for each material are given by
\begin{equation}
C=\rho\frac{g_{an}^2}{4\pi}\left(\frac{Z}{\upmu}+\frac{N}{\upmu}\right).
\label{eq22}
\end{equation}

In \eqnref{eq22}, $\rho$ is the density of the respective material already indicated above, $Z$ and $N$ are the number of protons and
the mean number of neutrons in the atoms of materials with mean  mass $M$ and $\upmu=M/m_{\rm H}$. For materials under consideration
$Z/\upmu=0.40422$, 0.46518, 0.50238, and 0.503205 for Au, Cr, Si, and SiO$_2$, respectively \cite{Fischbach:1999}.
The values of $N/\upmu$ are 0.60378, 0.54379, 0.50628, and 0.505179 for the same respective materials \cite{Fischbach:1999}.

The gradient of the pressure due to two-axion exchange is obtained from \eqnref{eq20} by the differentiation with
respect to $a$
\begin{equation}
P_{an}^{\prime}(a)=\frac{m_a\hbar^2}{m^2m_{\rm H}^2}\int_{1}^{\infty}du
\frac{\sqrt{u^2-1}}{u}\Psi\left(\frac{m_acu}{\hbar}\right)
\exp\,\left(-\frac{2m_acau}{\hbar}\right).
\label{eq23}
\end{equation}

The constraints on the parameters of axion-nucleon interaction, $g_{an}$, $m_a$, are obtained from the inequalities
\begin{equation}
\left| P_{an}(a)\right|<\Delta P(a), \qquad
P_{an}^{\prime}(a)<\Delta P^{\prime}(a),
\label{eq24}
\end{equation}
\noindent
similar to \eqnref{eq18}.

\begin{figure}[!h]
\centering
\includegraphics[scale=0.93]{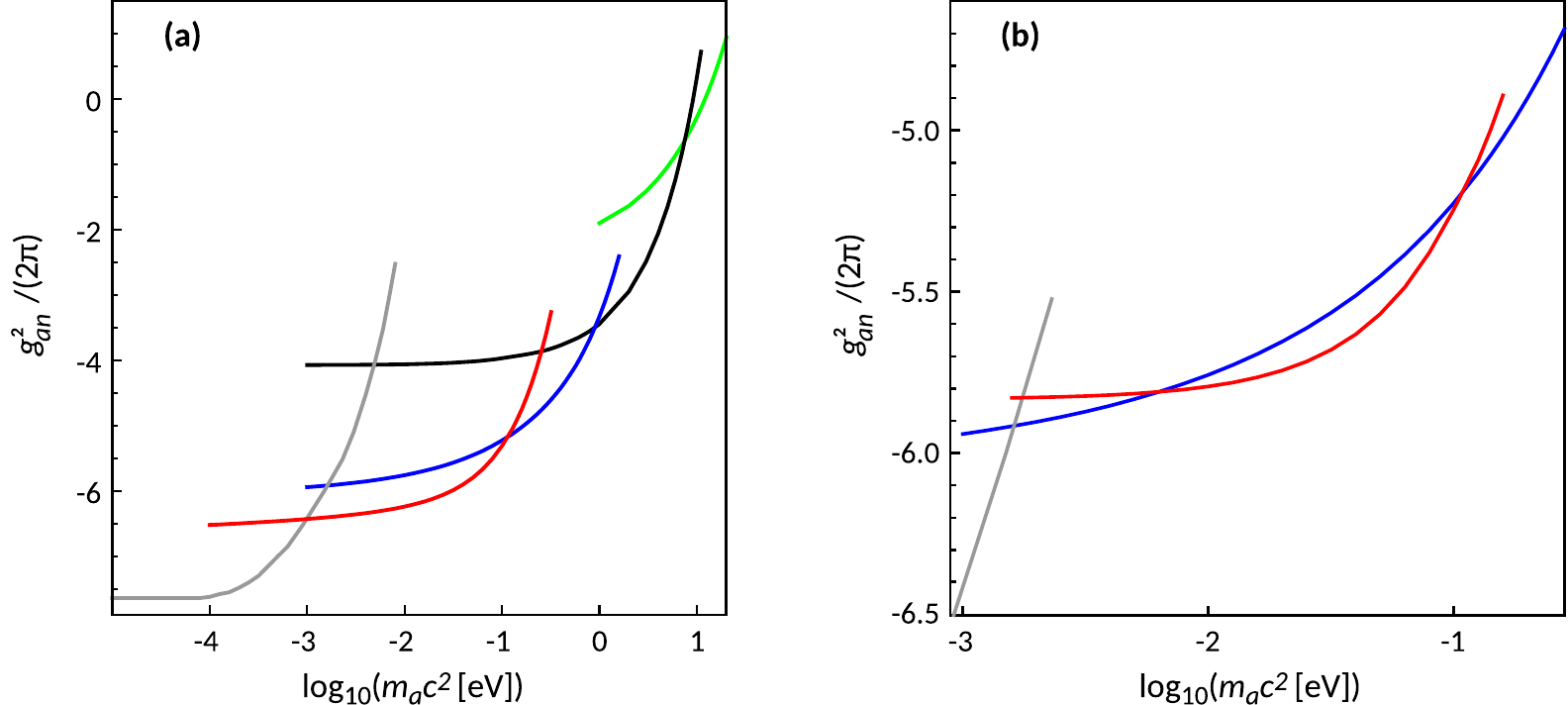}
\caption{\label{fgC2}Constraints on the axion-to-nucleon coupling constant obtained from the improved
Casimir-less experiment, the Cavendish-type experiment, measurements of the effective Casimir pressure, and the lateral Casimir force are shown by the blue, gray, black, and green lines, respectively. The proposed constraints which could be obtained from the \cannex{} test of the quantum vacuum when measuring (\textbf{a}) the Casimir pressure and (\textbf{b}) its gradient are indicated by the red lines.}
\end{figure}

To find the strongest prospective constraints on the axion-to-nucleon interaction, we have used the same parameters of the improved setup, as listed above when considering the Yukawa interaction. The computational results for $g_{an}^2/(2\pi)$ as a function of $m_ac^2$ obtained by using the first and second inequalities in \eqnref{eq24} are shown by the red lines in Figure~\ref{fgC2}a,b, respectively. Again, the space above each line is excluded (will be excluded, in the case of the red line) by the results of the respective experiment.

The strongest laboratory constraints obtained up to date \cite{Klimchitskaya:2015a} at the same region of axion masses are shown by the blue line.
They follow from the improved Casimir-less experiment \cite{Chen:2016}. For comparison purposes, the constraints following from the Cavendish-type experiment \cite{Kapner:2007,Adelberger:2007},
measurements of the effective Casimir pressure \cite{Bezerra:2014b,Decca:2007a,Decca:2007}, and the lateral Casimir force between corrugated surfaces~\mbox{\cite{Bezerra:2014c,Chiu:2009,Chiu:2010}} are shown by the gray, black, and green lines, respectively.
As can be seen  in Figure \ref{fgC2}a,b, the stronger constraints again follow from measuring the Casimir pressure. According to \figref{fgC2}a, the \cannex{} test of the quantum vacuum could strengthen the presently known constraints by up to a factor of 3.

\section{Discussion}
\label{sec:discussion}
In the foregoing, we have considered the \cannex{} test of the quantum vacuum intended to measure the gradient of the Casimir pressure between two parallel plates at separations exceeding a few micrometers.
As already discussed in the literature \cite{Almasi:2015a,Sedmik:2018}, this experiment would be capable to place stronger constraints on the chameleon model of dark energy and, thus, bring new information concerning the
origin of vacuum energy and the value of the cosmological constant.
Here, we proposed a modification of the \cannex{} setup, which allows for simultaneous measurements of both the pressure and its gradient. We also considered several improvements in the already existing setup which will allow for a more precise determination of several parameters. According to our results, with these improvements the \cannex{} test of the quantum vacuum could be used to directly measure for the first time the thermal effects in the Casimir pressure and its gradient at separations from 5 to 10 micrometers. We have also shown that this experiment could strengthen the presently known best constraints on the parameters of non-Newtonian gravity by up to a factor of 10 over a wide interaction range.
The~constraints on the axion-to-nucleon coupling constant could be strengthened by up to a factor of 3 in the region of axion masses from 1 to $100\,$meV.

\section{Conclusions}
\label{sec:conclusion}
To conclude, \cannex{} can be considered as a promising new laboratory experiment for investigating unusual features of the quantum vacuum, thermal effects in the Casimir forces,
non-Newtonian gravity and properties of axion-like particles being hypothetical constituents of dark matter. Together with other laboratory experiments, such as the demonstration of the dynamical Casimir effect in
superconducting circuits \cite{60,61},
the large-scale accelerator projects,
and astrophysical data, it may serve as a useful tool in resolving fundamental problems of the quantum vacuum.
\vspace{6pt}

\paragraph{Author Contributions: }Investigation, G.L.K., V.M.M., R.I.P.S., and H.A.;
writing, G.L.K., V.M.M., and R.I.P.S.;  project administration, H.A.

\paragraph{Funding: }This research was partially funded by the Russian Foundation for Basic Research grant number
19-02-00453 A.

\paragraph{Acknowledgments: }V.M.M.\ was partially supported by the Russian Government Program of Competitive Growth of Kazan Federal University. G.L.K. and V.M.M. are grateful to the Atominstitut of the TU Wien, where this work was performed, for kind hospitality. R.I.P.S. was supported by the TU Wien.

\end{document}